\documentclass[reprint,amsmath,amssymbaps,pre,twocolumn,showpacs,superscriptaddress]{revtex4-1}
\usepackage{latexsym}
\usepackage{amssymb}
\usepackage{graphicx}
\usepackage{amsmath}
\usepackage{bm}
\usepackage[colorlinks,
          linkcolor=black,
            citecolor=black,
            urlcolor=blue
           ]{hyperref}
\usepackage{verbatim}
\usepackage{slashed}
\usepackage{mathrsfs}
\usepackage{extarrows}
\usepackage{mathtools}
\usepackage{comment}
\usepackage{mathtools,slashed}
\usepackage{color}
\newcommand{\nc}{\newcommand}
\nc{\rnc}{\renewcommand}
\nc{\nn}{\nonumber}
\nc{\del}{{\partial}}
\rnc{\Im}{{\rm{Im}\,}}
\rnc{\Re}{{\rm{Re}\,}}
\nc{\db}{\displaybreak[0]\\}
\nc{\bra}{\langle}
\nc{\ket}{\rangle}

\nc{\lam}{\lambda}
\nc{\g}{{\mathfrak{g}}}
\nc{\zb}{\bar{z}}
\nc{\hb}{\bar{h}}
\nc{\J}{\mathcal{J}}
\nc{\su}{\widehat{\mathfrak{su}}(2)_k}

\nc{\tcr}{red}

\def\nn{\nonumber}
\def\beq{\begin{equation}}
\def\eeq{\end{equation}}
\def\bea{\begin{eqnarray}}
\def\eea{\end{eqnarray}}

\numberwithin{lemma}{section}
\numberwithin{proposition}{section}
\numberwithin{theorem}{section}
\numberwithin{corollary}{section}
\numberwithin{conjecture}{section}

\begin{document}

\title{Open spin chain realization of topological defect on 1d Ising model and boundary and bulk symmetry}

\author{Yoshiki Fukusumi}

\affiliation{ Zagreb University, faculty of science, Bijenička cesta 30 10000 Zagreb}
\author{Shumpei Iino}
\affiliation{Institute for Solid State Physics, University of Tokyo, Kashiwa, Chiba 277-8581, Japan}

\begin{abstract}
We study the realizations of topological defects in 1d quantum Ising model with open boundary
condition at criticality.
Applying the construction discussed in~[M. Hauru, G. Evenbly, W. W. Ho, D. Gaiotto, and G. Vidal, Phys. Rev. B {\bf 94}, 115125 (2016)],
we prove that the Ising model on an open chain with multiple
topological defects can be transformed to the same model with boundary magnetic fields and noninteracting
boundary degrees of freedom.
This results in the appearance of linear
combination of Cardy states~[J. L. Cardy, Nucl. Phys. B {\bf 324}, 581 (1989)], which
can be interpreted as an edge state of the spin or fermion chain.
We show that this edge state with the large boundary entropy can be protected under
bulk perturbation whereas it is fragile to a boundary perturbation.
Our formulation suggests an existence of nontrivial edge physics under the existence of topological
defects and opens many interesting questions for future analysis related to boundary and bulk physics.

\end{abstract}

\maketitle


\section{Introduction}

Massless field theories have played an essential role in the description of gapless systems~\cite{QFT_ZinnJustin}. Notably, the conformal field theory (CFT) is one of the most powerful and successful methods to study them, especially in $(1+1)$-dimensional critical systems~\cite{Belavin1984}. One celebrated example is the Tomonaga-Luttinger liquid (TLL), which describes the property of a large class of one-dimensional interacting electrons~\cite{10.1143/ptp/5.4.544,doi:10.1063/1.1704046}. {If electrons are physically confined to 1D, as
in a quantum wire, they can no longer be described as effectively
noninteracting. In other words, the Fermi liquid picture has to be
replaced by a TLL.}  The TLL can be explained as a special case of the $c=1$ CFT~\cite{GINSPARG1988153}, which also describes the critical system of the Gaussian universality appearing in a wide range of physics~\cite{giamarchi2004quantum}.

Considering boundaries or defects in gapless systems is one of the most important topics in CFTs~\cite{Cardy1984}. Such a system can be analyzed by boundary CFTs (BCFTs). BCFTs have been the significant theoretical framework to investigate not only boundary critical phenomena but also a wide variety of problems in condensed matter, high energy and mathematical physics~\cite{CARDY1989581,PhysRevB.46.15233, 1999tald.conf..473S, 2001JHEP...11..016G}. One important feature of BCFT is that not only the bulk but also the boundaries must satisfy the conformal invariance. Such a strong restriction results in the complete classification of the permissible boundary conditions for some simple cases~\cite{CARDY1989581,Behrend2000}, while it has been a difficult problem to classify them for the general BCFT~\cite{2002JHEP...06..028Q}.

{One of the most important applications of BCFTs in condensed matter
physics is the analysis of gapless systems where effectively
noninteracting electrons are coupled to a magnetic impurity via a spin
exchange in an s-wave channel, known as Kondo problem
\cite{10.1143/PTP.32.37}. This problem can be reduced to a 1D problem which is formally equivalent to quantum wire junction problem \cite{PhysRevB.46.15233,PhysRevLett.68.1220}, and the
corresponding BCFT successfully reveals the critical properties of
such a system \cite{Affleck:1991yq, Affleck:1990by,Affleck:1990iv}; for a
review see Ref. \cite{Affleck:1995ge}. The BCFT approach can be
used to describe the spin-exchange coupling to a spinful impurity \cite{PhysRevLett.75.300,PhysRevB.53.3211}, confirming earlier
renormalization-group results \cite{PhysRevLett.72.892}. As for the different
problem of potential scattering against a spinless impurity in a TLL,
it was shown that such an impurity can have significant effects on the
property of the conductivity of the electrons \cite{PhysRevLett.68.1220,PhysRevB.47.4631,PhysRevLett.75.2196}. Such a system
can be mapped onto a system with a boundary by folding it across the
impurity site \cite{Wong:1994np}.}

After this folding trick, one can consider the impurity problem as the boundary interaction problem of the folded model in general. Hence it becomes possible to analyze the system by the BCFT by the folding trick.
More generally, multiple impurities or multiple wire junctions problems have been investigated enthusiastically and many nontrivial boundary critical behaviors have been proposed~\cite{Affleck:1991yq,Oshikawa_2006,2015JHEP...07..072K}. Also, it is proposed that these settings could be realized in experiments in the junctions of quantum wires or quantum Hall edges, which can be theoretically studied in the framework of BCFTs~\cite{Affleck:1995ge,1999tald.conf..473S,PhysRevLett.89.075505,PhysRevLett.77.4612}. Moreover, junctions of 1d quantum systems can serve as a building block of quantum circuits~\cite{PhysRevB.59.15694}. In order to realize such models in the experiment and develop their application to mesoscopic physics, further investigations of the impurity problem have been still significant.

In BCFT, especially when a single impurity is repulsive, it divides the system into decoupled wires under the renormalization group (RG) flow~\cite{Affleck:1995ge,1999tald.conf..473S,PhysRevLett.89.075505,PhysRevLett.77.4612}. Such a repulsive impurity or defect should be treated as \textit{factorising defects}, which is known as a class of conformal defect~\cite{2007JHEP...04..095Q}. Since the cut ends of the decoupled regions at the defect correspond to the conformal boundary states, it is important to study the conformally invariant boundary conditions in order to understand this kind of defects further~\cite{2007JHEP...04..095Q, 2004JHEP...04..019G,2000NuPhB.588..552R,2000hep.th...10082G}.

Besides the factorising defects, another type of defects is known, called \textit{topological defects}~\cite{PETKOVA2001157}. A topological defect is a transmissive defect and can be thought of as an operator which generates a general ``twist" corresponding to the duality or the symmetry of the system~\cite{2001PhLB..517..429C}.
{The constructions of topological defects in the lattice models have been expolored by using integrability~\cite{Aasen:2016dop} and more recently by using tensor network~\cite{Lootens2019}.} {Also, the RG behavior of topological defect has been studied, for example, in \cite{Chang:2018iay,2004JHEP...04..019G}.}
 
More generally, the construction and classification of various conformal defects, which may connect different CFTs, have been discussed~\cite{2007JHEP...04..095Q,2014NuPhB.885..266K,2015JHEP...07..072K,2012JHEP...12..103G, CRNKOVIC1990637}. In particular, the defects between two CFTs connected by an integrable bulk perturbation are algebraically constructed~\cite{2012JHEP...12..103G, CRNKOVIC1990637}. Surprisingly, this algebraic construction is shown to be consistent with the perturbative calculation of the identity defect~\cite{Brunner_2016}. Hence further research is required about the relations between RG flows of CFT and (purely) algebraic construction of conformal defects in order to grasp the nature of phase transitions.

Since these two types of defects, the factorising and topological defects, can be defined independently, in principle one could consider the systems where both types of defects coexist. It is practically significant to consider the multiple-defect systems with these two types, because real materials can have the topological defects in the presence of open boundaries, which can be described as a factorising defect.

Most of the existing works, nevertheless, are considering the system with either topological or factorising defects. Hence, as a simplest example, we study the 1d transverse field Ising model~\cite{PFEUTY197079} on the lattice in the presence of both open boundary conditions and topological defects. We show that it can be transformed into the same model without any topological defects by using the property of the defects. In this construction, we demonstrate that the appearance of the linear combination of the conformally invariant boundary states~\cite{2004JHEP...04..019G} can be interpreted as the edge degrees of freedom, which is similar to the recently advocated gapless topological phases~\cite{PhysRevX.7.041048,2019arXiv190506969V}. Moreover, our result also suggests the nondecreasing of boundary entropy $g$~\cite{PhysRevLett.67.161} under the \textit{bulk} perturbation even in the lattice model, which may signal a nontrivial edge physics~\cite{2006hep.th....9034F, 2008NuPhB.798..491G, 2009JPhA...42W5403F}.

We would like to comment that the 1d transverse field Ising model is not only a toy model easy to analyze theoretically but also can be realized in the experiment~\cite{HEID1995123,PhysRevB.60.3331,MAARTENSE197793,Coldea2010}. Therefore we would expect that our results could be observed experimentally in the future, after the realization and precise control of the topological and factorising defects in the experiments.

The remain of this paper is structured as follows. In Sec.~\ref{sec:bcft}, we introduce the topological defects and the conformal boundary states more precisely than in this section, and review the basic concepts of the BCFT, especially the \textit{Cardy states}. We also explain the known properties of the Ising BCFT in $1+1$ dimension. In Sec.~\ref{sec:Ising}, we discuss the transverse field Ising chain on the lattice. After introducing the lattice realization of the Cardy states and the topological defects, it is elaborated on how the topological defects interact with the open boundaries. We demonstrate that the topological defects in the bulk can be absorbed into the boundary states, where the fusion between the defects and the boundaries obeys the fusion rule of the Ising CFT. We also show that the fusion with the multiple topological defects yields the superposition of the conformally invariant boundary states at the boundary. The extension to the fermionic Ising BCFT is also discussed. After discussing the RG argument of the boundary states in Sec.~\ref{sec:RG}, we conclude our results in Sec.~\ref{sec:conclusion}.

\section{Topological defect and its action to boundary state\label{sec:bcft}}

\subsection{The topological defect and the factorising defect}

First of all, we begin with reviewing the \textit{conformal defects}, since both of the topological and factorising defects can be understood as a special class of the conformal defects. The conformal defects, which connect CFT${}_{1}$ and CFT${}_{2}$, are line objects $X$ which satisfy
\begin{equation}
  T_{1}-\overline{T_{1}}=T_{2}-\overline{T_{2}},
  \label{eq:conformal_defect}
\end{equation}
along the defect line, where $T_{i}$ is the energy momentum tensor of CFT${}_i$ and $\overline{T}_i$ is the antiholomorphic counterpart~\cite{PETKOVA2001157,2007JHEP...04..095Q}. The complete classification of the conformal defects for two different CFTs is a difficult problem and has never been accomplished. Actually, by folding the system along the defect, the problem reduces to the construction of boundary states in the product theory CFT${}_{1}\times$CFT${}_{2}$, which may in general break extended symmetry such as Lie group symmetry~\cite{2002JHEP...06..028Q, 2007JHEP...04..095Q, 2014NuPhB.885..266K, 2015JHEP...07..072K}. Though the construction of these boundary states is a fascinating and difficult problem, in this paper we concentrate on connecting the same CFTs (i.e., CFT${}_1=$ CFT${}_2$) by two classes of conformal defects which are called topological defect and factorising defect, especially for $A$ series minimal CFT with diagonal torus partition function~\cite{CAPPELLI1987445}.

The topological defects for CFTs were extensively studied by Zuber and Petkova~\cite{PETKOVA2001157}. They are transmissive conformal defects, which can be moved and deformed continuously without changing the behavior of the system~\cite{2007JHEP...04..095Q}. The topological defects satisfy the stronger condition than in Eq.~(\ref{eq:conformal_defect}): $T_{1}=T_{2}$ and $\overline{T_{1}}=\overline{T_{2}}$, along the defect line.

For Ising model, there exist three topological defects $D_{\boldsymbol 1}$, $D_{\epsilon}$, $D_{\sigma}$ where each index of the defects corresponds to the primary operator with the conformal dimension $h_{\boldsymbol 1}=0$, $h_{\epsilon}=\frac{1}{2}$, $h_{\sigma}=\frac{1}{16}$, respectively~\cite{CARDY1986200,1996PhRvL..77.2604O,PETKOVA2001157}. These defects satisfy the following fusion rule,
\begin{align}
  D_{\epsilon}\times D_{\epsilon}&=D_{\boldsymbol 1}, \nonumber\\
  \label{eq:defect_fusion}
  D_{\epsilon}\times D_{\sigma}&=D_{\epsilon}, \\
  D_{\sigma}\times D_{\sigma}&=D_{\boldsymbol 1}+D_{\epsilon}\nonumber.
\end{align}
It should be noted that, although this fusion rule is the same as that of bulk fields of the Ising CFT, the correspondence between the bulk operator and the topological defect is not always true.

On the other hand, a factorizing defect is a totally repulsive conformal defect. More specifically, it is described by Dirichlet boundary condition~\cite{2004JHEP...04..019G}, which can be expressed as the operator
\begin{equation}
  X=\sum_{a,b} f_{a,b} |a\ket \bra b|,
\end{equation}
where $|a\ket$ and $|b\ket$ are the conformal boundary states which we will define precisely in Sec.~\ref{subsec:bcft} and $f_{a,b}$ is a coefficient.

Our strategy for constructing topological defects in the presence of the open boundary conditions in 1d spin chain is quite simple:
\begin{enumerate}
\item We consider the realization of topological defects on a spin chain with the periodic boundary condition.
\item Then we insert a factorising defect which is equivalent to assigning open boundary condition.
\item Finally, We put it near the open boundary and make the defect absorbed into the boundary. We can move the topological defect freely by some unitary transformation thanks to its transmissiveness.
\end{enumerate}
In this paper, we only consider a specific lattice model, 1d quantum Ising model at criticality because this third step requires some detailed information of the spin chain~\cite{PhysRevB.94.115125}. However, we believe the above construction can be applied to other general models described by CFT because the existence of topological defects could be assumed under the symmetries or the dualities of the system~\cite{2007NuPhB.763..354F,FUCHS2011673}.

\subsection{BCFT and Cardy states\label{subsec:bcft}}

BCFT is defined by assigning boundary condition which can preserve conformal symmetry of the theory~\cite{Cardy1984}. Even for the general theory described by the extended algebra larger than the Virasoro algebra, one can consider symmetry breaking or preserving boundary conditions which can preserve conformal symmetry~\cite{2002JHEP...06..028Q}.

Here, we concentrate on the established case, the $A$ series minimal model with the diagonal partition function~\cite{CARDY1989581}. The conformal boundary state satisfies the following boundary condition,
\begin{equation}
  \left(L_{n}-\overline{L}_{-n} \right)|B\ket =0,
\end{equation}
where $L_{n}$ is the generator of the local conformal transformation and $\bar{L}_n$ is the antiholomorphic one. A solution of this equation is given by the linear combination of the Ishibashi states,
\begin{equation}
  |j\ket\ket =\sum_{M} |j, M\ket \otimes \overline{|j, M \ket},
  \label{eq:Ishibashi}
\end{equation}
where $j$ is an index of the primary fields which contain the identity operator labeled by $0$, and $M$ labels its descendant level.

However, the Ishibashi state is not a ``physical" basis if we consider the open string partition function $\bra\bra j| e^{-\tau H_{\mathrm{CFT}}}|k \ket\ket=\delta_{j,k}\chi_{j}(\tau)$. By using the modular $S$ transformation, the annulus partition function should be written as,
\begin{equation}
 Z_{j,k}(\tau)= \bra j| e^{-\tau H_{\mathrm{CFT}}} |k  \ket =\sum_{i} n^{i}_{j,k}\chi_{i}\left(-1/\tau\right),
\end{equation}
where $n^{i}$ is a nonnegative integer matrix and $|j\ket$ is some boundary state labeled by the index $j$. As an easiest solution, Cardy obtained $|B_{a}\ket =\sum_{j}\frac{S_{aj}}{\sqrt{S_{0j}}}|j\ket\ket$ where he has taken $n$ as fusion matrix $N$~\cite{CARDY1989581}. This solution of the conformally invariant boundary state is called the \textit{Cardy state}.

For the Ising BCFT, it is know that there exist three different Cardy states labeled as $|{\boldsymbol 1}\ket$, $|\epsilon\ket$, and $|\sigma\ket$, respectively~\cite{CARDY1989581}. Similarly to the topological defects, these boundary states are related to the three primary operators in the Ising CFT. Notice that, in the language of the transverse field Ising chain, these Cardy states correspond to the fixed points of the boundary states in the sense of the boundary RG:
\begin{equation}
  |{\boldsymbol 1}\ket=|+\ket,\ |\epsilon\ket=|-\ket,\ |\sigma\ket=|\mathrm{free}\ket,
  \label{eq:Ising_Cardy}
\end{equation}
where $|\mathrm{free}\ket$ corresponds to the disordered boundary state with the $Z_2$ symmetry while $|\pm\ket$ represents the magnetized and $Z_2$ symmetry-broken boundary state with $+$ or $-$ spin, respectively.

We would like to comment that when one considers lattice models with boundary it may be natural to start with the boundary perturbation theory to Cardy states in BCFT~\cite{2000hep.th...10082G}. In this setup, some nontrivial existence of boundary states described by the linear combination of the Cardy states is shown by this boundary RG argument of the least relevant boundary field $\psi_{(1,3)}$ in the minimal CFT~\cite{2000hep.th...10082G, 2000NuPhB.588..552R, 2013JPhA...46n5401K}. Actually, the complete description of boundary perturbation theory inevitably needs extensive analytical calculations such as RG analysis and truncated conformal space approach (TCSA).

\subsection{Graham-Watts states}

By considering multiplication of the topological defect to the Cardy states, one can obtain another series of ``physical" states which are represented by the linear combination of the Cardy states~\cite{2004JHEP...04..019G},
\begin{equation}
  D_{a}| B_{b}\ket =\sum_{j}N^{j}_{ab} |B_{j}\ket =|a\times b\ket.
\end{equation}
We call these states \textit{Graham-Watts} states.

The RG flow from a Cardy state $|B_{a}\ket$ to another one $|B_{b}\ket$ implies the existence of RG flow from a Graham-Watts state $D_{d}|B_{a}\ket$ to $D_{d}|B_{b}\ket$ for an arbitrary index $d$~\cite{2004JHEP...04..019G}. Hence the Graham-Watts state can give an easy construction of the extended boundary states with the information of RG flow. The essential part of their proof of the existence of such a flow is quite simple. First, we think about two annulus partition functions $Z_{d, a}$ and $Z_{d, b}$. The RG flow from $Z_{d, a}$ to $Z_{d, b}$ is implied by the RG flow from $|B_{a}\ket$ to $|B_{b}\ket$. Second we interpret the boundary state $|B_{d}\ket$ as $D_{d}|0\ket$. Because the topological defect $D_{d}$ can freely move, we can obtain the RG flow from $Z_{0,d\times a}$ to $Z_{0,d\times b}$, which suggests the boundary RG flow of the Graham-Watts states from $D_{d}|B_{a}\ket$ to $D_{d}|B_{b}\ket$.

For the Ising model, by multiplying $D_{\epsilon}$ and $D_{\sigma}$ recursively to Cardy states, we can obtain the following set of states,
\begin{equation}
  |+\ket , \ |-\ket , \ 2^{n} | \mathrm{free} \ket , \ 2^{n} \left( | +\ket +|- \ket \right) ,
\end{equation}
where $n$ is a positive integer including $0$. One can see these results from the fact that $D_{\epsilon}$ just flips the spin in the boundary as $D_{\epsilon}|\pm\ket=|\mp\ket$ and $D_{\epsilon}|\mathrm{free}\ket=|\mathrm{free}\ket$, while $D_{\sigma}$ makes the $Z_2$-symmetric boundaries as $D_{\sigma}|\pm\ket=|\mathrm{free}\ket$ and $D_{\sigma}|\mathrm{free}\ket=|+\ket+|-\ket$ (remember the fusion rule of the Ising CFT like in Eq.~(\ref{eq:defect_fusion}) and the definition of the Ising Cardy states in Eq.~(\ref{eq:Ising_Cardy})). In this setup, the states like $|+\ket +|-\ket +|\mathrm{free} \ket$ do not appear.

The lattice realization of the cat state $ | +\ket +|- \ket $ was first considered in the BCFT analysis of the tricritical Ising model as far as we know~\cite{Chim1996, Affleck2000, 2004JSMTE..03..001F}, and recently this state in the Ising model started to capture an attention in the field of the symmetry-protected topological phases~\cite{PhysRevX.7.041048, 2019arXiv190506969V}. We believe, at least with respect to the bulk and boundary RG and the topological defect theory, that the appearance of these Graham-Watts state is ubiquitous.

We would like to comment that if one thinks about the (boundary and bulk) RG flow~\cite{2006hep.th....9034F,2008NuPhB.798..491G,2015JHEP...12..114K}, the Graham-Watts states like  $|+\ket+|-\ket$, $2|\text{free}\ket$ exhibit the coincidentally similar behavior to the edge state of the SPT or the intrinsic topological phases~\cite{PhysRevX.7.041048,2019arXiv190506969V}, as will be discussed in Sec.~\ref{sec:RG}.

{It should also be noted that there exists more general lattice models, A-D-E lattice models, which should be described by minimal conformal field theories \cite{Behrend2001}. In these models, topological defects and open boundary conditions can be implemented by using the technique of (boundary) Yang-Baxter equation and these models can be mapped to anyonic chains \cite{Feiguin:2006ydp,Buican:2017rxc}. However, in these works, what boundary operator may trigger the boundary RG flow is not fully classified, especially for Granham-Watts states.
Hence by applying our argument of the boundary (and bulk) RG flow of Graham-Watts states, one can possibly predicts or even construct nontrivial edge modes in the open anyonic chain in general. These RG analysis only requires transformation law of boundary operators by applying topological defects and the RG flows of Cardy's states as we have discussed.}

\section{Ising model with topological defects and open boundary condition\label{sec:Ising}}



\subsection{Lattice realization of topological defects on the critical Ising chain\label{subsec:revIsinglattice}}

We consider the quantum Ising model on a semi-infinite chain with the following Hamiltonian:
\beq
\label{eq:IsingHam}
H=-\left[\sum_{i=1}^{\infty}\left(\sigma_i^z\sigma_{i+1}^z+\Gamma\sigma_i^x\right)+h\sigma_1^z\right],
\eeq
where $\sigma^{\alpha}$ with $\alpha=x,y,z$ are the Pauli matrices, $\Gamma$ is the transverse field, and $h$ is the longitudinal field only on the boundary. Tuning $\Gamma=1$ brings this model to the gapless point, whose low-energy physics can be described by the $(1+1)$-dimensional Ising CFT.

As is already mentioned, corresponding to the three primary fields, there can be three conformally-invariant boundary conditions in the Ising CFT~\cite{CARDY1989581}: $|{\boldsymbol 1}\ket = |+\ket$, $|{\epsilon}\ket = |-\ket$, $|{\sigma}\ket = |\mathrm{free}\ket$, each of which represents the fixed boundary condition with $+$ spin and $-$ spin in the $\sigma^{z}$ basis, and the free boundary condition, respectively. These boundary conditions can be realized by controlling the boundary magnetic field $h$ in Eq.~(\ref{eq:IsingHam}). $h=0$ yields the free boundary condition, which has a relevant field with the scaling dimension $1/2$ since the conformal spectra of this boundary fixed point are ${\boldsymbol 1}\oplus \epsilon$. Because this relevant field corresponds to the boundary external field $h$, the infinitesimal magnetic field $h$ induces the flow to the ordered boundary states at $h=\pm\infty$: $h>0$ corresponds to $|+\ket$ state while $h<0$ to $|-\ket$ state~\cite{2013JPhCS.411a2004B}.

The topological defects in the Ising CFT can also be realized on the lattice by controlling the parameters in the Hamiltonian~\cite{PhysRevB.94.115125}. Corresponding to the primary fields, the classified three topological defects in the Ising CFT are $D_{\boldsymbol 1}$, $D_{\epsilon}$, and $D_{\sigma}$~\cite{2002hep.th....9048G}, as is also introduced in Sec.~\ref{sec:bcft}. The $D_{\boldsymbol 1}$ represents the trivial $Z_2$ symmetry defect, which means there is no defect and has no effect in the system. The $D_{\epsilon}$ is called the (nontrivial) $Z_2$ symmetry defect, whose lattice realization is the same as the antiperiodic boundary condition. Therefore, the insertion of the symmetry defect into the bond between the $i$-th and $(i+1)$-th sites on the critical Ising chain can be performed by the following transformation for the parameter in the Hamiltonian:
\beq
\label{eq:transfSym}
\sigma_i^z\sigma_{i+1}^z \rightarrow -\sigma_i^z\sigma_{i+1}^z,
\eeq
which means a change of the interaction between the $i$-th and $(i+1)$-th spin into an antiferromagnetic one. We refer to this situation as `there is a $D_{\epsilon}$ in the $(i,i+1)$-bond'.

The last defect $D_{\sigma}$ is called the Kramers-Wannier (KW) duality defect, which can be inserted into the $i$-th site on the lattice by~\cite{2002hep.th....9048G, PhysRevB.94.115125}
\beq
\label{eq:transfDual}
\sigma_{i-1}^z\sigma_i^z + \sigma^x_i \rightarrow \sigma_{i-1}^z\sigma_i^y.
\eeq

\subsection{The fusion between the Cardy states and topological defects}

Now we have the lattice realization of the Cardy states and the topological defects on the critical Ising chain. In this subsection, we consider the fusion between them on the lattice. The key point in our analysis is that the topological defects can be moved freely by using the appropriate unitary transformations~\cite{2002hep.th....9048G}. Namely, after the insertion of the topological defects $D_a$ into the bulk, we can move it near the boundary and finally have it absorbed into the edge states $|b\ket$ by the unitary transformations. The resulting boundary states should be, according to the conjecture of CFT, consistent with the fusion rule of the Ising CFT, $|a\times b\ket$. We demonstrate that the boundary states on the gapless Ising chain obtained by being fused with the topological defects are consistent with the conjecture of the Ising CFT\cite{PETKOVA2001157}.

\subsubsection{the symmetry defects}

First of all, we discuss the trivial $Z_2$ symmetry defect $D_{\boldsymbol 1}$. Since the lattice realization of $D_{\boldsymbol 1}$ is just the absence of any defects, trivially the unitary transformation for moving it is just an identity transformation. Then the boundary states remain unchanged after the fusion with the $D_{\boldsymbol 1}$, which is consistent with the fusion rules of the identity operator and arbitrary operators in the Ising CFT:
\bea
{\boldsymbol 1}\times {\boldsymbol 1} &=& {\boldsymbol 1}, \\
{\boldsymbol 1}\times {\epsilon} &=& \epsilon, \\
{\boldsymbol 1}\times {\sigma} &=& \sigma.
\eea

Next, we discuss the fusion of the Cardy states and the nontrivial symmetry defect $D_{\epsilon}$. We consider the critical Ising model on a semi-infinite chain with the symmetry defect in the $(2,3)$-bond:
\beq
\label{eq:IsingHamSym}
H_{D_{\epsilon}}^{(2,3)}\equiv -\left[\sigma_1^z\sigma_2^z-\sigma_2^z\sigma_{3}^z+\sum_{i=3}^{\infty}\sigma_i^z\sigma_{i+1}^z+\sum_{i=1}^{\infty}\sigma_i^x+h\sigma_1^z\right],
\eeq
which can be obtained by Eq.~(\ref{eq:transfSym}). A unitary transformation to move the defect from the $(2,3)$-bond into the $(1,2)$-bond is the Pauli matrix $\sigma^x_2$:
\beq
\begin{split}
\sigma^x_2H_{D_{\epsilon}}^{(2,3)}{\sigma^x_2}^{\dagger}&=-\left[-\sigma_1^z\sigma_{2}^z+\sum_{i=2}^{\infty}\sigma_i^z\sigma_{i+1}^z+\sum_{i=1}^{\infty}\sigma_i^x+h\sigma_1^z\right] \\
&\equiv H_{D_{\epsilon}}^{(1,2)}.
\end{split}
\eeq

Just in the same way, we can move the defect from $(1,2)$-bond to the boundary: i.e., let it absorbed into the boundary by $\sigma_1^x$ transformation:
\beq
\sigma_1^xH_{D_{\epsilon}}^{(1,2)}{\sigma_1^x}^{\dagger}=-\left[\sum_{i=1}^{\infty}\sigma_i^z\sigma_{i+1}^z+\sum_{i=1}^{\infty}\sigma_i^x-h\sigma_1^z\right],
\eeq
which is equivalent to the Hamiltonian without any defect Eq.~(\ref{eq:IsingHam}) whose boundary external field is flipped. Therefore, the fusion of $D_{\epsilon}$ have the following effects on the Cardy states:
\bea
D_{\epsilon}|+\ket &=& |-\ket, \\
D_{\epsilon}|-\ket &=& |+\ket, \\
D_{\epsilon}|\mathrm{free}\ket &=& |\mathrm{free}\ket.
\eea
Notice that these relations are consistent with the fusion rules between $\epsilon$ operator and the primary fields corresponding to each Cardy state:
\bea
\epsilon\times {\boldsymbol 1} &=& \epsilon, \\
\epsilon\times {\epsilon} &=& {\boldsymbol 1}, \\
\epsilon\times {\sigma} &=& \sigma.
\eea
Now we are able to confirm that the effect of the $D_{\epsilon}$ on the boundary states is consistent with the Ising CFT.

\subsubsection{the duality defect}

We turn to the discussion of the duality defect $D_{\sigma}$. Let us consider the critical Ising chain with $D_{\sigma}$ at the third site, whose Hamiltonian can be obtained by Eq.~(\ref{eq:transfDual}):
\beq
\begin{split}
  H_{D_{\sigma}}^{(3)} \equiv &-\left(\sigma_1^z\sigma_{2}^z+\sigma_2^z\sigma_3^y
  +\sum_{i=3}^{\infty}\sigma_i^z\sigma_{i+1}^z\right) \\
  &-\left(\sigma_1^x+\sigma_2^x+\sum_{i=4}^{\infty}\sigma_i^x\right)-h\sigma_1^z.
\end{split}
\eeq
The unitary transformation which transfers the defect into the second site can be defined as
\beq
U_{3\rightarrow 2}=\left[\left(R_y^{\frac{\pi}{4}}R_x^{\frac{\pi}{4}}\right)_2\otimes\left(R_z^{\frac{\pi}{4}}\right)_3\right]\otimes CZ_{2,3},
\eeq
where $\left(R_a^{\theta}\right)_i=\cos\theta\times{\boldsymbol 1}_i+i\sin\theta\times\sigma_i^a$ and
\beq
CZ_{i,i+1}=\left(|\uparrow\ket\bra\uparrow|\right)_{i+1}{\boldsymbol 1}_i+\left(|\downarrow\ket\bra\downarrow|\right)_{i+1}\sigma_i^z
\eeq
is the control Z operator~\cite{PhysRevB.94.115125}. The simple calculation actually results in
\beq
\begin{split}
&U_{3\rightarrow 2}H_{D_{\sigma}}^{(3)}{U_{3\rightarrow 2}}^{\dagger} \\
&=-\left[\left(\sigma_1^z\sigma_{2}^y +\sum_{i=2}^{\infty}\sigma_i^z\sigma_{i+1}^z\right)+\left(\sigma_1^x+\sum_{i=3}^{\infty}\sigma_i^x\right)+h\sigma_1^z\right]\equiv H_{D_{\sigma}}^{(2)},
\end{split}
\label{eq:IsingHamDualtmp}
\eeq
which represents the Hamiltonian with $D_{\sigma}$ at the second site.

Applying the unitary transformation $U_{2\rightarrow 1}$ for Eq.~(\ref{eq:IsingHamDualtmp}) yields the absorption of the defect into the boundary. The resulting Hamiltonian is
\beq
\begin{split}
U_{2\rightarrow 1}H_{D_{\sigma}}^{(2)}{U_{2\rightarrow 1}}^{\dagger}&=-\left[\sum_{i=1}^{\infty}\sigma_i^z\sigma_{i+1}^z+\sum_{i=2}^{\infty}\sigma_i^x+h\sigma_1^y\right] \\
&\equiv H_{D_{\sigma}}^{(1)},
\end{split}
\label{eq:IsingHamDualRes}
\eeq
where the boundary field is applied along the $y$-direction and the boundary transverse field term $\sigma_1^x$ is absent.

The boundary states of Eq.~(\ref{eq:IsingHamDualRes}) can be interpreted as follows. When the original boundary state before the fusion with $D_{\sigma}$ is in the free boundary condition, there is no boundary longitudinal field, $h=0$. Since in this case the Hamiltonian (\ref{eq:IsingHamDualRes}) is commutable with $\sigma_1^z$, the ground state $|\psi\ket$ can be decomposed into two different sectors depending on the parity of $\bra\psi|\sigma_1^z|\psi\ket=\pm 1$. For each $\pm$ sector, the Eq.~(\ref{eq:IsingHamDualRes}) can be described as
\beq
H_{D_{\sigma}}^{(1)}=-\left[\pm\sigma_{2}^z+\sum_{i=2}^{\infty}\sigma_i^z\sigma_{i+1}^z+\sum_{i=2}^{\infty}\sigma_i^x\right],
\label{eq:IsingHamDualDeg}
\eeq
which is equivalent to the critical Ising Hamiltonian (\ref{eq:IsingHam}) with $h=\pm 1$ and without any defect. Since any finite boundary longitudinal field $h$ induces the ordered boundary states, the resulting boundary state for each parity sector is $|\pm\ket$, respectively. Therefore, because the boundary state of Eq.~(\ref{eq:IsingHamDualRes}) is the superposition of the boundary states of the two parity sectors, we can conclude that
\beq
D_{\sigma}|\mathrm{free}\ket=|+\ket+|-\ket.
\label{eq:DualCardy1}
\eeq

When the original boundary states before applying $D_{\sigma}$ is in the fixed boundary conditions $|\pm\ket$, on the other hand, the Hamiltonian Eq.~(\ref{eq:IsingHamDualRes}) is no longer commutable with $\sigma_1^z$ due to $h\neq 0$. Here let us focus on the case where $h>0$ and the original boundary state is $|+\ket$. As is already explained, since the fixed point for the free boundary condition is unstable for the perturbation of $h$, a positive finite $h$ flows to $h=\infty$ for the boundary RG, which allows us to analyze the Hamiltonian Eq.~(\ref{eq:IsingHamDualRes}) with the limit of $h\rightarrow \infty$ taken. Therefore the ground state $|\psi\ket$ of Eq.~(\ref{eq:IsingHamDualRes}) necessarily maximize $h\sigma_1^y$, which amounts to
\beq
|\psi\ket = \frac{1}{\sqrt{2}}|\sigma^{y}=1\ket_1\otimes |s_2,s_3,\cdots\ket,
\label{eq:IsingDualWF}
\eeq
where the spin state at the boundary is determined as the eigenstate of $\sigma_1^y$ with the eigenvalue $+1$ and the other part is described as $|s_2,s_3,\cdots\ket$. Notice that Eq.~(\ref{eq:IsingDualWF}) yields $\bra\psi|\sigma_1^y|\psi\ket=1$ and $\bra\psi|\sigma_1^z|\psi\ket=0$. Therefore the effective Hamiltonian results in
\beq
H_{D_{\sigma}}^{(1)}=-\left[\sum_{i=2}^{\infty}\sigma_i^z\sigma_{i+1}^z+\sum_{i=2}^{\infty}\sigma_i^x+h\right],
\eeq
which is equivalent to the critical Ising Hamiltonian with the free boundary condition, since the constant term $h$ has only irrelevant effect.

Although we discuss only the case of $h>0$, the negative $h$ also yields the essentially same effective Hamiltonian,
\beq
H_{D_{\sigma}}^{(1)}=-\left[\sum_{i=2}^{\infty}\sigma_i^z\sigma_{i+1}^z+\sum_{i=2}^{\infty}\sigma_i^x-h\right],
\eeq
the difference of which from the case of $h>0$ is just the constant term. As a result, we can conclude
\beq
D_{\sigma}|\mathrm{+}\ket=D_{\sigma}|\mathrm{-}\ket=|\mathrm{free}\ket.
\label{eq:DualCardy2}
\eeq

In conclusion, according to Eq.~(\ref{eq:DualCardy1}) and Eq.~(\ref{eq:DualCardy2}), we are able to observe that the fusion between the duality defect and the Cardy states are consistent with the fusion rules of the Ising CFT:
\bea
\sigma\times {\boldsymbol 1} &=& \sigma, \\
\sigma\times {\epsilon} &=& \sigma, \\
\sigma\times {\sigma} &=& {\boldsymbol 1}+\epsilon.
\eea

\subsubsection{comment on the stability of $|+\ket+|-\ket$}

As shown in Eq.~(\ref{eq:DualCardy1}), the boundary state generated by the fusion with the free boundary and the duality defect is the superposition of the Cardy states, $|+\ket+|-\ket$, which can be realized in the Hamiltonian Eq.~(\ref{eq:IsingHamDualRes}) with $h=0$. As is explained in Eq.~(\ref{eq:IsingHam}), changing the transverse field $\Gamma$ from the critical value $1$ induces the gapped ground state. Let us consider perturbing Eq.~(\ref{eq:IsingHamDualRes}) with $h=0$ into off-criticality by controlling the transverse field:
\beq
H_{D_{\sigma}}^{(1)}=-\left[\sum_{i=1}^{\infty}\sigma_i^z\sigma_{i+1}^z+\Gamma\sum_{i=2}^{\infty}\sigma_i^x\right].
\eeq
Because even for the non-critical system the Hamiltonian commutes with $\sigma_1^z$, the ground state can be decomposed  again into the $\pm$ parity sectors, each of which Hamiltonians is
\beq
H_{D_{\sigma}}^{(1)}=-\left[\pm\sigma_2^z+\sum_{i=2}^{\infty}\sigma_i^z\sigma_{i+1}^z+\Gamma\sum_{i=2}^{\infty}\sigma_i^x\right].
\eeq
Even if the bulk is gapped, the boundary orders for a finite boundary longitudinal field, then the boundary state of this Hamiltonian is also the superposition of two ordered boundary states fixed with $+$ spins and $-$ spins. This suggests that the boundary state at criticality $|+\ket+|-\ket$ is stable against the bulk perturbation breaking the KW self-dual symmetry. We discuss later this in the viewpoint of boundary RG flow.

\subsection{fusion with multiple defects}

We have confirmed on the lattice that the fusion between the topological defects $D_a$ and the Cardy states $|b\ket$ can be derived by the fusion rule in the Ising CFT, which yields the resulting boundary states $|a\times b\ket$. Just in the same way, we can prove that fusing another defect $D_c$ with this obtained boundary state $|a\times b\ket$ on the critical Ising chain yields the boundary state $|a\times b\times c\ket$.

As an example, let us consider multiplying the duality defect twice for the Cardy states in the Ising CFT. The Hamiltonian whose boundary takes in a single $D_{\sigma}$ is Eq.~(\ref{eq:IsingHamDualRes}). For this Hamiltonian, we insert the other duality defect into the second site:
\beq
H_{D_{\sigma}}^{(1),(2)}\equiv -\left[\sigma_1^z\sigma_{2}^y+\sum_{i=2}^{\infty}\sigma_i^z\sigma_{i+1}^z+\sum_{i=3}^{\infty}\sigma_i^x+h\sigma_1^y\right].
\eeq
Then we move the inserted defect into the boundary by the appropriate unitary transformation:
\beq
U_{2\rightarrow 1}H_{D_{\sigma}}^{(1),(2)}{U_{2\rightarrow 1}}^{\dagger}= -\left[\sum_{i=2}^{\infty}\sigma_i^z\sigma_{i+1}^z+\sum_{i=2}^{\infty}\sigma_i^x+h\sigma_1^x\sigma_2^z\right].
\label{eq:IsingHamDualTwice}
\eeq

Since $\sigma_1^x$ commutes with the Hamiltonian, the boundary states are the superpositions of the one for each $Z_2$ parity sector. When $h\neq 0$, the Hamiltonian for these two $\pm$ sectors are
\beq
U_{2\rightarrow 1}H_{D_{\sigma}}^{(1),(2)}{U_{2\rightarrow 1}}^{\dagger}= -\left[\sum_{i=2}^{\infty}\sigma_i^z\sigma_{i+1}^z+\sum_{i=2}^{\infty}\sigma_i^x\pm h\sigma_2^z\right],
\eeq
which are equivalent to the critical Ising chain with a boundary longitudinal field $\pm h$, respectively. This means $D_{\sigma}D_{\sigma}|+\ket = D_{\sigma}D_{\sigma}|-\ket = |+\ket+|-\ket$, which is consistent with the fusion rule of $\sigma\times\sigma\times{\boldsymbol 1}=\sigma\times\sigma\times\epsilon={\boldsymbol 1}+\epsilon$. For $h=0$, on the other hand, the boundary site $1$ does not interact with any other sites. This represents the existence of boundary degrees of freedom which is equivalent to the two-fold degenerated free boundary states $2|\mathrm{free}\ket$. Hence it is possible to understand the prefactor "$2$" of the Graham-Watts state as a boundary degree of freedom. Notice that these results are consistent with the fusion rule of $\sigma\times\sigma\times\sigma=2\sigma$.

Now we see the fusion of the Cardy states with multiple duality defects on the gapless Ising chain yields the same results with the Ising CFT. The other case of multiple applications of the topological defects can also be easily proved. By considering the insertion of topological defects recursively, one can obtain the boundary states like $2^{n}\left( |+\rangle +|-\rangle \right)$ and $2^{n}| \text{free}\rangle$, where the factor $2^{n}$ can be interpreted as boundary degree of freedom which does not interact with the bulk.

\subsection{duality connection between Ising and fermionic Ising BCFT}

In this section, we introduce the possible relation of boundary states between Ising CFT and fermionic CFT~\cite{2020arXiv200105055R,2020arXiv200212283H}. In 1d lattice model, these two models are realized by Ising spin chain and Kitaev chain, which are related by Jordan-Wigner transformation~\cite{Kitaev_2001}. Because of the nonlocality of this transformation, the meaning of symmetries can be changed in these two representations~\cite{2014AnPhy.351.1026G, 2013PhRvB..87d1105C, 2010PhRvL.104b0402C}. For example, the topologically protected edge degree of freedom of Kitaev chain can be understood as the symmetry protected degree of freedom in the spin chain which prohibits the existence of the boundary disorder field~\cite{2019arXiv190506969V}.

First of all, the Cardy states of Ising CFT are written as~\cite{CARDY1989581},
\begin{align} 
  |+\ket &=\frac{1}{\sqrt{2}}|\boldsymbol 1\ket\ket+\frac{1}{\sqrt{2}}|\epsilon\ket\ket +\frac{1}{\sqrt{2}\sqrt[4]{2}}|\sigma\ket\ket, \\
  |- \ket&=\frac{1}{\sqrt{2}}|\boldsymbol 1\ket\ket+\frac{1}{\sqrt{2}}|\epsilon\ket\ket -\frac{1}{\sqrt{2}\sqrt[4]{2}}|\sigma\ket\ket ,\\
  |\text{free} \ket&=|\boldsymbol 1\ket\ket-|\epsilon\ket\ket,
\end{align}
where the double cap states $|j\ket\ket$ are the Ishibashi states introduced in Eq.~(\ref{eq:Ishibashi}). The application of the $Z_{2}$ spin flip, which is expressed as $|\sigma\ket\ket=-|\sigma\ket\ket$, to these states results in the following transformations: $|+\ket \rightarrow |-\ket$, $|-\ket \rightarrow |+\ket$, and $|\mathrm{free}\ket \rightarrow |\mathrm{free}\ket$. Hence the total Hilbert space, spanned by the positive integer linear combination of Cardy states, does not change by this transformation.

However, if we think about the $Z_{2}$ transformation $|\epsilon\ket\ket \rightarrow -|\epsilon\ket\ket$, which corresponds to high and low temperature dual transformation \cite{2019PhR8271S,2019ScPP77K}, to the Cardy states, the total Hilbert space is no longer the same as the original one. For example, the state $|+\ket +|-\ket$ is transformed into $\sqrt{2}|\text{free} \ket$ and it is not an integer multiplication of Cardy states. Hence we have to think about this transformation and the resultant Hilbert space in BCFT other than Ising BCFT. Interestingly, the Hilbert space spanned by the transformed boundary states by this dual transformation coincides with that of the fermionic Ising BCFT, which is recently proposed~\cite{2020arXiv200105055R, 2020arXiv200212283H}. In the following we note the detail of this correspondence.

First, we decompose the Ising BCFT basis to the symmetric and antisymmetric sector under $Z_{2}$ spin flip transformation. The symmetric sector is spanned by the two basis $|+\ket+|-\ket$ and $|\text{free}\ket $, whereas the antisymmetric sector is spanned by $\pm\left( |+\ket-|-\ket\right)$. Then by applying the  $Z_{2}$ transformation $|\epsilon\ket\ket \rightarrow -|\epsilon\ket\ket$, we can obtain the following relations,
\begin{align}
  |\mathrm{free}\ket &\rightarrow |\text{fixed},+\ket_{\mathrm{NS}}= |\text{fixed},-\ket_{\mathrm{NS}}  \\
  |+\ket+|-\ket&\rightarrow  |\text{free}\ket_{\mathrm{NS}},\\
  \pm\left(|+\ket-|-\ket \right)&\rightarrow |\pm,\text{fixed}\ket_{\mathrm{R}}
\end{align}
where the right hand side represents the boundary states in the fermionic CFT in~\cite{2020arXiv200105055R} and NS and R represent Neveu-Schwartz and Ramond sector.

Hence, if we think about the connection between Kitaev chain and Ising chain which are connected by Jordan-Wigner transformation, it may be natural to guess the former is $Z_{2}$ dual of the latter~\cite{2019PhR8271S, 2019ScPP77K}. As global (bulk and boundary) $Z_{2}$ spin flip does not change the partition function or energy spectrum, the global duality transformation may also not change the partition function or energy spectrum. Because this duality relates low temperature physics and high temperature physics, there may exist close relation between massless (massive) flow of Ising CFT and massive (massless) flow of fermionic Ising CFT.

\section{RG arguments of our model}
\label{sec:RG}

\subsection{$g$-factor and boundary degree of freedom}

In this section, we discuss the behavior of Graham-Watts state on the Ising chain under bulk and boundary interaction and its implication on RG argument of BCFT.  First, we introduce the general aspects of the $g$-factor~\cite{PhysRevLett.67.161} for the Graham-Watts states with respect to the boundary and bulk RG flow. In general, beginning with the Cardy state characterized by the identity index, the $g$-factor or boundary entropy takes the following form~\cite{2019arXiv191106041K},
\begin{equation}  
  g_{a\times b}=g_{a}\frac{g_{b}}{g_{0}}.
\end{equation}
In minimal CFT, it was also pointed out that the insertion of a topological defect inevitably increases the $g$-factor. Hence we can say that the Graham-Watts states have more edge degrees of freedom than the original Cardy states before the multiplication of the defect. Moreover, when we think about boundary $g$-theorem, protection of these boundary degrees of freedom may need more symmetry which can exclude relevant boundary perturbation~\cite{PhysRevLett.67.161,2004PhRvL..93c0402F}.

While we have commented on the boundary perturbation on the Graham-Watts state with respect to $g$-factor, we also would like to discuss the effect of the \textit{bulk} perturbation on the Graham-Watts state. The most important thing to note is that the $g$-factor can increase under bulk perturbation~\cite{2015JHEP...12..114K, 2008NuPhB.798..491G}.

In other words, it means a boundary degree of freedom can be protected (or even can be enhanced) by the bulk perturbation. Hence we can expect that some Graham-Watts states unstable against the boundary perturbation (or protected by boundary symmetry) may survive under bulk renormalization. As we will show in the next subsections, the state $|+\ket +|-\ket$ and $2|f\ket$ are the first two examples of this case in the lattice transverse field Ising model. For further research, it should be stressed that some Cardy states can flow to a Graham-Watts state by bulk perturbation~\cite{2000hep.th...10082G, 2000NuPhB.588..552R, 2013JPhA...46n5401K}. This phenomenon can be related to the appearance of the edge states in topological ordered phases, although it has never been explained as such a consequence of RG flow as far as we know. Considering $c$-theorem and $g$-theorem of CFT~\cite{1986JETPL..43..730Z, PhysRevLett.67.161}, this phenomenon is unusual because it shows a nondecreasing of the degree of freedom under RG flow.

\subsection{RG flow of Graham-Watts state $|+\ket +|-\ket$}

As one can see, in Ising model the non-Cardy state $|+\ket +|-\ket$ can be easily realized. Here we note a boundary and bulk RG argument of this state in the framework of the minimal CFT~\cite{2009JPhA...42W5403F}.

First, as we have shown in the previous subsections, breaking boundary KW duality induces the flow from this state to $|\text{free}\ket$, which is triggered by the boundary disorder field $\mu$. This result is consistent with the boundary RG flow of Graham-Watts states derived from the boundary RG flow of Cardy states from $|\text{free}\ket$ to $|+\ket$ induced by the boundary order operator $\sigma$~\cite{2019arXiv191106041K}.

Second, the bulk perturbation maintains $|+\ket +|-\ket$ in the sense that the twofold degeneracy on the boundary is robust against the bulk KW duality breaking. Actually, it is also consistent with massless flow of the minimal model with boundary. In some literatures~\cite{2009JPhA...42W5403F, 2015JHEP...12..114K}, the following boundary RG flows are discussed: $|+\ket =|I\ket_{\text{Ising}}\rightarrow|I\ket_{M(2,3)} $ and $|-\ket =|\phi_{1,3}\ket_{\text{Ising}}\rightarrow |\phi_{1,2}\ket_{M(2,3)}$. Therefore, considering each sector of Ising chain Hamiltonian which corresponds to boundary spin value $\sigma^{z}=\pm1$, we can achieve the flow from $|+\ket +|-\ket$ to the state $|I\ket_{M(2,3)} +|\phi_{1,2}\ket_{M(2,3)}$ under boundary RG induced by the bulk perturbation. Hence we can see the conservation of boundary degrees of freedom.

However, we would like to note that there exist some subtleties of the state $|I\ket +|\phi_{1,2}\ket$. Actually, we have not used the identification of Kac formula for $M(2,3)$ model. $M(2,3)$ is known as trivial CFT with bulk fields with conformal dimension $0$, but there may still exist nontrivial boundary critical phenomena, known as percolation. Moreover, $\phi_{1,2}$ field and its singular vector are known to be described by Schramm-Loewner evolution~\cite{1992JPhA...25L.201C, SMIRNOV2001239}. (There exists similar problem for the identification $\phi_{2,1}=\phi_{1,3}$ for Ising model~\cite{PhysRevLett.119.191601}.) Hence we will suggest that it may be an open problem whether we can use the relation $\phi_{1,2}\sim I$ at the boundary. If this identification is true, one can observe exact 4-fold degeneracy for finite spin chain (but it is unlikely to happen as we will discuss in the next subsection). As can be seen our lattice Hamiltonian, there should also exist similar preservation of boundary degree of freedom under massive flow.

Finally, we mention a general argument of the duality defect~\cite{2007NuPhB.763..354F}. Duality defect $D_{d}$ is a defect which implements the symmetry $g\in G$ as,
\begin{equation}
  D_{d}\times D_{d}=\sum_{g\in G}D_{g}.
\end{equation}
where $D_{g}$ is the symmetry defect of the theory.

The minimal CFT $M(m,m+1)$ has the $Z_{2}$ symmetry, generated by the primary operator $\phi_{1,m}$~\cite{2019arXiv191106041K}. The $Z_{2}$ defect labeled by this index changes the primary field as follows,
\begin{equation} 
  D_{1,m}\phi_{r,s}=\phi_{m-r,s}.
\end{equation}
Hence, in this model, there may exist duality defects with the condition,
\begin{equation}
  D_{d}\times D_{d}=I+D_{1,m}.
\end{equation}
For example, Ising model has a duality defect $D_{1,2}=D_{\sigma}$ and tricritical Ising model has a duality defect, $D_{2,1}=D_{\sigma'}$. Hence when we interpret this relation by using Grham-Watts states, there may exist similar phenomena in various models.

\subsection{Flow from $|+\ket+|-\ket$ to $|\mathrm{free}\ket$ and bulk massless flow}

As we have shown, the boundary state $|+\ket+|-\ket$ is robust against bulk perturbation. Moreover, by using Jordan-Wigner transformation, it explains the topologically protected edge state of Kitaev chain\cite{2014AnPhy.351.1026G, 2013PhRvB..87d1105C, 2010PhRvL.104b0402C}.

The point is that there exists boundary flow from $|+\ket+|-\ket$ to $|\mathrm{free}\ket$ and $|+\ket+|-\ket$ has the larger $g$-factor. In this section, we will review what may happen if we added bulk perturbation to this flow. Let us introduce the following lattice Hamiltonian,
\begin{equation}
  H=-\sum_{i=2}^{\infty}\left(\sigma_{i}^z\sigma^{z}_{i+1}+\Gamma\sigma^{x}_{i}\right)-\sigma^{z}_{1}\sigma^{z}_{2}+h_{x}\sigma^{x}_{1}.
\end{equation}
$h_{x}=0$ corresponds to the boundary state $|+\ket+|-\ket$ at criticality and it flows to $|\mathrm{free}\ket$ by choosing $h_{x}\neq 0$. In the massless flow, because of the spontaneous symmetry breaking, we can conclude $h_{x}\rightarrow 0$.

Hence the system is described by the flow of $|+\ket+|-\ket$, which has the larger $g$-factor. In the massive flow, this phase is the disordered phase and the spin chain is decoupled which is characterized by the eigenvalue of $\sigma^{x}$ of each site. Hence the boundary perturbation is still relevant and the system is described by the flow of $|\mathrm{free}\ket$.

\subsection{Flow from $2|\mathrm{free}\ket$ to $|+\ket+|-\ket$ and bulk perturbation}

In this  subsection, we consider the boundary flow from  $2|f\ket$ to $|+\ket+|-\ket$ and its behavior induced by bulk perturbation. For this purpose, we consider the following Hamiltonian,
\begin{equation}
  H=-\sum_{i=2}^{\infty}\left(\sigma_{i}^z\sigma^{z}_{i+1}+\Gamma\sigma^{x}_{i}\right)+h\sigma^{z}_{1}\sigma^{z}_{2}.
\end{equation}
By choosing $\Gamma=1$ and $h=0$, the boundary condition is described by $2|\mathrm{free}\ket$, as we have already shown. This state has the larger $g$-factor than $|+\ket+|-\ket$, and it flows to this state by boundary interaction $h\neq 0$ at criticality.

In fact, this boundary state is robust against bulk perturbation. To see this, we consider the situation off criticality. In the massless flow, the effect of $h\neq 0$ cannot be negligible by spontaneous symmetry breaking. Hence the boundary flow should become that of $|+\ket +|-\ket$. In the massive phase, the interaction becomes irrelevant because the system is in the disordered phase. Hence the boundary spin $\frac{1}{2}$ degree freedom can survive in this regime and it is similar to an edge state of the Haldane phase.

Our model is trivial to some extent, but it coincidentally shows similar behavior to the phase transition between the Haldane phase and the ferromagnetic phase, recently considered in~\cite{2019arXiv190506969V}. This phase transition is protected by boundary symmetry which prohibits the boundary interaction $\sigma^{z}_{1}$ and $\sigma^{x}_{1}$ and bulk $Z_{2}$ spin flip symmetry. This boundary symmetry may result from the original spin $1$ XXZ Heisenberg chain.

\section{conclusion\label{sec:conclusion}}

We have discussed the realization of topological defects in the presence of open boundaries, using the 1d transverse field Ising model on the lattice. We have analytically shown that the model can be transformed into the same model without any defect but with boundary fields and boundary degrees of freedom. Especially, it has been demonstrated that one can understand the appearance of the edge states such as $|+\ket +|-\ket$ or $2|\mathrm{free}\ket$, as a consequence of the application of the duality defects. Compared with other types of defects or impurities, the characteristics of topological defects that they can move smoothly and satisfy fusion algebra simplify the multiple defects problem.

More generally, we expect that our formulation may suggest the existence of general boundary states $D_{a}|B\ket$ on a 1d critical spin chain as an edge mode. This structure is similar to the edge state of the SPT phases as we have discussed in Sec.~\ref{sec:RG}. Hence the boundary and bulk RG argument of CFT might be useful for a unified explanation of topological phases. In our analysis, it should be noted that the appearance of degenerate edge mode is a result from the fusion of the topological defects and Cardy states. For a more complete analysis of the edge state, one has to consider the boundary RG flow from BCFT to boundary topological quantum field theory starting from boundary states with the large $g$-values.

As a related problem, it is interesting to consider the general realization of the flow with this nondecreasing $g$-factor in general lattice models. The present paper treats the case which preserves the boundary degree of freedom under bulk RG flow, but the existing paper may also predict increasing of the boundary degree of freedom under RG flow~\cite{2008NuPhB.798..491G, 2015JHEP...12..114K}. It may generate an emergent boundary degree of freedom with an emergent boundary symmetry under the bulk RG. Hence it might be possible to predict that the bulk RG flow with boundaries can explain the appearance of nontrivial edge modes of gapped systems.
{For this purpose, it may be important to consider anyonic chains with topological defects and open boundary which should corresponds to minimal CFT with topological defects and open boundary condition. By studying the behaviors of these lattice models follwing our discussion, one can learn fundamental aspects of RG flows of Graham-Watts states as appearance of notrivial edge modes.}

Finally, it should be noted that some theoretical applications of ("smeared") BCFT to periodic gapped systems which inevitably contain the linear combination of Cardy states are recently proposed by Cardy~\cite{SciPostPhys.3.2.011}. More recently, his conjecture was checked by using TCSA for some specific models~\cite{2019JHEP...01..177L}. Hence further analysis of general boundary states as we have constructed in this work may shed new light on the analysis of the RG flow to the gapped system and its realization in the lattice models~\cite{Ares_2020}.


\section*{Acknowledgements}
YF thanks Ryohei Kobayashi for introducing the paper~\cite{2019arXiv190506969V}, and related discussion.
YF also thanks to the previous fruitful collaboration with Yuan Yao, which is closely related
to this project.
YF also thanks Osor Bari\v si\' c  for his careful reading of the manuscript and some discussion of many-body-localization and its
relation to the local integral of motion. 
We thank Zohar Nussinov, Jacques H.H. Perk, and Ryan Thorngren  for helpful comments and discussions.
YF acknowledges the support by the QuantiXLie Center of Excellence, a project co-financed by the Croatian Government and European Union through the European Regional Development Fund - the Competitiveness and Cohesion Operational Programme (Grant KK.01.1.1.01.0004).
SI is grateful to the support of Program for Leading Graduate Schools (ALPS).

\appendix
\section{Kramers-Wannier transformation for the finite size open spin chain}
In this section, we review the KW transformation on the finite-length Ising model with open boundary condition~\cite{PhysRev.162.436,2001NuPhB.596..513W}. The point is that the Hamiltonian is NOT invariant when one considers the boundary condition, but the partition function is invariant under this transformation. Moreover, this transformation can be thought of as the shift of the duality defect to one boundary to the other.

First, we introduce the following Hamiltonian of the Ising model,
\begin{equation}
H=-\sum_{i=1}^{N-1}\left[ \sigma^{z}_{i}\sigma^{z}_{i+1}+\sigma^{x}_{i+1}\right] +h_{1}\sigma^{x}_{1}+h_{N}\sigma^{z}_{N},
\end{equation}
where the left boundary can be represented by $|+\ket +|-\ket =D_{\sigma}|\text{free}\ket$ with $h_{1}=0$, and $|\text{free}\ket=D_{\sigma}|+\ket=D_{\sigma}|-\ket$ otherwise. Notice that the left boundary states can be represented as the product of the duality defect and a Cardy state as $D_{\sigma}|B\ket$. The right boundary, on the other hand, can be represented by $|\text{free}\ket$ with $h_{N}=0$, and $|+\ket$ or $|-\ket$ with $h_{N}\neq 0$.

Next, we introduce the KW transformation~\cite{2010PhRvL.104b0402C},
\begin{align}
  \sigma^{z}_{i}\sigma^{z}_{i+1}=\sigma'^{x}_{i}. \\
  \prod_{j=1}^{i}\sigma^{x}_{j}=\sigma'^{z}_{i}.
\end{align}
with $\sigma^{z}_{N}=\sigma'^{x}_{N}$, and $i=1, ...,N$.

After this transformation, the Hamiltonian becomes,
\begin{equation}
  H'=-\sum_{i=1}^{N-1}\left[ \sigma'^{z}_{i}\sigma'^{z}_{i+1}+\sigma'^{x}_{i}\right] +h_{1}\sigma'^{z}_{1}+h_{N}\sigma'^{x}_{N}.
\end{equation}
As one can easily see, the boundary terms are not invariant. After this transformation, the duality defect in the left boundary condition has been eliminated as $D_{\sigma}|B_{\text{left}}\ket\rightarrow |B_{\text{left}}\ket$, and the right boundary condition acquires a duality defect as $|B_{\text{right}}\ket\rightarrow D_{\sigma}|B_{\text{right}}\ket$. Hence we can understand the KW transformation as moving the duality defect from one boundary to the other.

In fact, this shows the lattice realization of Graham-Watts argument when we consider the left and right edges\cite{2004JHEP...04..019G}. For example, one can see that the $\sigma^{z}_{N}$ which induces the boundary RG flow from $|\text{free}\ket$ to $|\pm\ket$ becomes $\sigma'^{x}_{N}$ which induces the boundary RG flow from $|+\ket+|-\ket$ to $|\text{free}\ket$. In the lattice models, such an argument may be trivial, but it is far from trivial if one considers the boundary RG argument of BCFT. Hence it may be helpful to consider a lattice model as verification of BCFT analysis such as TCSA. It should be noted that, as we have mentioned in the main text, there exist some subtleties of the operator identification by using Kac table when one considers boundary fields.

As we have discussed, the duality transformation does not preserve boundary states. Hence we state here more precise implication of this transformation, with respect to the correlation functions. First, we introduce the $n$ point correlation function of the disorder operators $\sigma'^{z}_{i_1}$, $\sigma'^{z}_{i_2}$, $\cdots$, $\sigma'^{z}_{i_n}$ as
\begin{equation}
  \bra D_{\sigma}B_{\text{left}}| \prod_{j}^{n}\sigma_{i_{j}}'^{z}|B_{\text{right}}\ket.
\end{equation}
By using KW transformation, this multipoint correlation function is equivalent to
the following correlation function of order operators,
\begin{equation}
  \bra B_{\text{left}}| \prod_{j}^{n}\sigma_{i_{j}}^{z}D_{\sigma}|B_{\text{right}}\ket.
\end{equation}
It should be noted that the boundary condition with $|\pm\ket$ for both boundaries is outside of this equivalence because we cannot move duality defect from one boundary to the other in this situation (The both boundaries cannot be represented as $D_{\sigma}|B\rangle$).

\section{Kramers-Wannier duality and dual $Z_{2}$ charge}

Here, we discuss the global $Z_{2}$ symmetry of the Ising model and its implication on the boundary RG argument. As we have stressed in the main text, such arguments require extensive calculation like TCSA if we investigated that kind of things in the rigorous sense. Hence we restrict our discussion only on the lattice model, from which we extract some implications on BCFT analysis.

First, we introduce the Hamiltonian with the following form (for simplicity we concentrate on left boundary indexed by the cite number $1$),
\begin{equation}
  H=-\sum_{i\ge 1}\left[ \sigma^{z}_{i}\sigma^{z}_{i+1}+\sigma^{x}_{i+1}\right] +h_{z}\sigma^{z}_{1}+h_{x}\sigma^{x}_{1}.
\end{equation}
When $h_{z}=0$, this Hamiltonian has $Z_{2}$ symmetry generated by the operator $\prod_{i\ge 1}\sigma^{x}_{i}$, and the boundary condition is represented by $|\text{free}\ket$, or $|+\ket+|-\ket$. In other words, the boundary operator $\sigma^{z}_{i}$ is charged under this symmetry.

As we have discussed, $\sigma^{x}_{1}$ triggers the boundary flow from  $|+\ket+|-\ket$ to $|\text{free}\ket$ which is closely related to the boundary flow from $|\text{free}\ket$ to $|\pm \ket$. Hence it may be natural to consider that $\sigma^{x}_{1}$ as charged object.

Actually this operator is charged under KW dual global $Z_{2}$ symmetry generated by the operator $\prod_{i}\sigma'^{x}_{i}=\sigma^{z}_{1}\sigma^{z}_{n}$ which only acts on the boundary of spin chain in the original representation. The same operator and related discussion can be found in \cite{2013PhRvB..87d1105C}.

As KW dual transformation implies, this boundary operator $\sigma^{x}_{1}$ induces the boundary flow from  $|+\ket+|-\ket$ to $|\text{free}\ket$, which corresponds to dual $Z_{2}$ symmetry breaking. As was stressed in \cite{2019arXiv190506969V}, $\sigma^{x}_{1}$ should be treated as boundary disorder operator which is different from order operator and energy operator with respect to $Z_{2}$ and dual $Z_{2}$ charge (in \cite{2019arXiv190506969V}, it was also shown that it is possible to eliminate the relevant boundary disorder operator if the boundary disorder operator is charged under some nontrivial symmetry transformations by assigning the symmetry on the total Hamiltonian. In these cases, the dual $Z_{2}$ symmetry can only be broken irrelevantly under these symmetries).

It should also be noted that this boundary disorder operator may have the conformal dimension $\frac{1}{2}$~\cite{2020arXiv200105055R}, which is different from conformal dimension $\frac{1}{16}$ of the order operator~\cite{2001NuPhB.596..513W}.

This may imply that the duality defect (and corresponding duality transformation) can change the conformal dimension of the boundary operators in general. Hence we believe that this type of argument in the lattice models, such as $Z_{N}$ parafermion model, suggests a lot of boundary RG flows which are quite nontrivial from the view of BCFT.

\section{Generalization to $Z_{N}$ Fateev-Zamolodchiknov model}

Our analysis in Sec.~\ref{sec:Ising} in the main text can be applied to $Z_{N}$ Fateev-Zamolodchikov model which can be described by $Z_{N}$ parafermion or some $c=1$ theory~\cite{Fateev:1985mm, 1994IJMPA...9.4921A} ( or complex CFT~\cite{2018JHEP...10..108G,PhysRevB.99.195130} ). The ferromagnetic Hamiltonian which corresponds to $Z_{N}$ parafermion can be written as,
\begin{align}
  H&=H_{\mathrm{bulk}}+H_{\mathrm{boundary}}, \\
  H_{\mathrm{bulk}}&=-\sum_{j=1}^{\infty}\sum_{k=1}^{N-1}\frac{1}{\text{sin}\frac{k\pi}{N}} \left(Z^{k}_{j}Z^{N-k}_{j+1}+X^{k}_{j} \right), \\
  H_{\mathrm{boundary}}&=-\sum_{k=1}^{N-1}\frac{1}{\text{sin}\frac{k\pi}{N}} h Z^{k}_{0}Z^{N-k}_{1}.
\end{align}
where $Z$ and $X$ satisfy by the following relations
\begin{align}
  Z^{N}&=X^{N}=1, \\
  Z^{\dagger}&=Z^{N-1}, \\
  X^{\dagger}&=X^{N-1}, \\
  ZX&=\omega XZ
\end{align}
with $\omega=e^{\frac{2i\pi}{N}}$ and $Z$ is diagonal matrix with eigenvalues $1$, $\omega$, ...., $\omega^{N-1}$.

$Z_{0}$ commutes with Hamiltonian and we can decompose the Hamiltonian by each sector corresponding to the eigenvalue of this boundary operator, $1, \omega ,..., \omega^{N-1}$. For example, in the sector with the eigenvalue $1$, we can obtain,
\begin{equation}
  H_{Z_{0}=1}=H_{\mathrm{bulk}}-\sum_{k=1}^{N-1}\frac{1}{\text{sin}\frac{k\pi}{N}}hZ^{k}_{1}.
\end{equation}
For the other eigenvalues, we can obtain the almost same expression and each expression can transform each other by the cyclic $Z_{N}$ transformation generated by $\prod_{j=1}^{\infty}X^{k}_{j}$. Hence we can conclude that this model has at least $N$ boundary state which is protected by dual $Z_{N}$ symmetry.

For finite-size spin chain, by considering unitary transformation to the left edge to assign eigenvalue $1$, we can obtain $N$ boundary states with degeneracy $N$ corresponding to the eigenvalue of the right edge. If the edge of the above model goes to some conformal boundary state $|B_{1}\ket$, the total state is described by applying $Z_{N}$ cyclic transformation $\Omega$ recursively, $\sum_{k=0}^{N-1} \Omega^{k}|B_{1}\ket$.
It seems natural to name this state as $Z_{N}$ duality state which is in close relation to duality defect with fusion $D_{d}\times D_{d}=\sum_{g\in Z_{N}}D_{g}$ \cite{2007NuPhB.763..354F} and its robustness under bulk perturbations. By applying parafermionic Jordan-Wigner transformation, one can observe similar protected edge state of this model as Ising and Kitaev chain~\cite{DOREY1996317}. Although further investigation of this model with boundary magnetic field is desired, it is out of the scope of this paper.

\bibliography{bibdefect.bib}
\end{document}